%% file: 20171117PRL_MT.tex
\documentclass[aps,prl,twocolumn,superscriptaddress,showkeys,showpacs]{revtex4-1}
\usepackage[T1]{fontenc}
\usepackage{standalone}
\usepackage{nccfoots}
\usepackage{comment}
\usepackage{graphicx}
\usepackage{braket}
\usepackage{xcolor}
\usepackage{placeins}
\usepackage{verbatim}
\usepackage{todonotes}
\usepackage{amssymb}
\usepackage{bbold}

\definecolor{mygreen}{rgb}{0,0.5,0}
\definecolor{myblue}{rgb}{0,0,0.75}
\definecolor{mymagenta}{cmyk}{0,1,0,0.12}
\usepackage[colorlinks=true,
   linkcolor=blue,
   citecolor=blue,
   urlcolor =blue]{hyperref}

\newcommand{\citeSM}{\cite[{\tiny SM}\kern-0.3em][]{SM}}

\newcommand{\be}{\begin{equation}}
\newcommand{\ee}{\end{equation}}

\newcommand{\tr}[2][]{\text{Tr}_{#1}\left[#2\right]}
\newcommand{\probs}[2][]{\ensuremath{\left\langle \text{P}\left(#2\right)^{#1} \right \rangle}}

\newcommand{\probsU}[2][]{\ensuremath{\text{P}\left(#2\right)^{#1} }}
\renewcommand{\vec}[1]{\mathbf{#1}}
\newcommand{\hilbertl}{\mathcal{N}_l}
\newcommand{\hilbertn}{\mathcal{N}_A}
\newcommand{\hilbert}{\mathcal{N}_A^{(N,S_z)}}
\newcommand{\ketbra}[2]{\ensuremath{\ket{#1}\bra{#2}}}

\expandafter\let\csname equation*\endcsname\relax

\expandafter\let\csname endequation*\endcsname\relax

\usepackage{amsmath}

\graphicspath{{Figures/}}

\begin{document}

\title{R{\'e}nyi Entropies from Random Quenches in Atomic Hubbard and Spin Models}

\author{A. Elben}

\thanks{These two authors contributed equally.}
\affiliation{Institute for Theoretical Physics, University of Innsbruck, Innsbruck,
Austria}
\affiliation{Institute for Quantum Optics and Quantum Information, Austrian Academy
of Sciences, Innsbruck, Austria}

\author{B. Vermersch}

\thanks{These two authors contributed equally.}
\affiliation{Institute for Theoretical Physics, University of Innsbruck, Innsbruck,
Austria}
\affiliation{Institute for Quantum Optics and Quantum Information, Austrian Academy
of Sciences, Innsbruck, Austria}

\author{M. Dalmonte}
\affiliation{Abdus Salam International Center for Theoretical Physics, 34151 Trieste, Italy}

\author{J. I. Cirac}
\affiliation{Max-Planck-Institut f\"ur Quantenoptik, Hans-Kopfermann-Str. 1, D-85748 Garching, Germany}

\author{P. Zoller}

\affiliation{Institute for Theoretical Physics, University of Innsbruck, Innsbruck,
Austria}
\affiliation{Institute for Quantum Optics and Quantum Information, Austrian Academy
of Sciences, Innsbruck, Austria}
\affiliation{Max-Planck-Institut f\"ur Quantenoptik, Hans-Kopfermann-Str. 1, D-85748 Garching, Germany}

\date{\today}
\begin{abstract}
We present a scheme for measuring R{\'e}nyi entropies in generic atomic Hubbard and spin models using single copies of a quantum state and for partitions in arbitrary spatial dimension.
Our approach is based on the generation of random unitaries from random quenches, implemented using engineered time-dependent disorder potentials, and standard projective measurements, as realized by quantum gas microscopes.
By analyzing the properties of the generated unitaries and the role of statistical errors, with respect to the size of the partition, we show that the protocol can be realized in exisiting AMO quantum simulators, and used to measure for instance area law scaling of entanglement in two-dimensional spin models or the entanglement growth in many-body localized systems.\end{abstract}
\maketitle

Atomic physics provides us with the realization of engineered quantum
many-body lattice models. This includes Hubbard models for bosonic
and fermionic cold atoms in optical lattices~\cite{Bloch2012}, and
spin models with Rydberg atoms~\cite{Browaeys2016} and chains of
trapped ions~\cite{Blatt2012}. Among the noticeable
recent experimental advances are quantum control, and single shot measurements
in lattice systems of atoms~\cite{Murmann2015,Haller2015,Greif2016,Boll2016,Cheuk2016,Labuhn2016,Zeiher2017,Bernien2017}
and ions~\cite{Jurcevic2017,Zhang2017} achieving \emph{single site
resolution}, as illustrated for atoms in optical lattices by the quantum
gas microscope~\cite{Kuhr2016}. This provides us not only with a
unique atomic toolbox to prepare equilibrium and non-equilibrium states
of quantum matter, but also with the opportunity to access in experiments
novel classes of observables, beyond the familiar low order correlation
functions. An outstanding example is the measurement of R{\'e}nyi entropies,
defined as $S^{(n)}(\rho_{A})=\frac{1}{1-n}\log{\rm Tr}(\rho_{A}^{n})$
($n>1$) with $\rho_{A}=\tr[\mathcal{S}\backslash A]{\rho}$ the reduced density matrix of a subsystem
$A\subset\mathcal{S}$ of a many-body system $\mathcal{S}$, which
gives us a unique signature of entanglement properties in many-body
phases and dynamics~\cite{Eisert2010}, and is also of interest in the ongoing discussion
on `quantum supremacy'~\cite{Boixo2016,BermejoVega2017,Gao2017,Bremner2017,Boixo2017}.

\begin{figure}[h!]
\includegraphics[width=0.97\columnwidth]{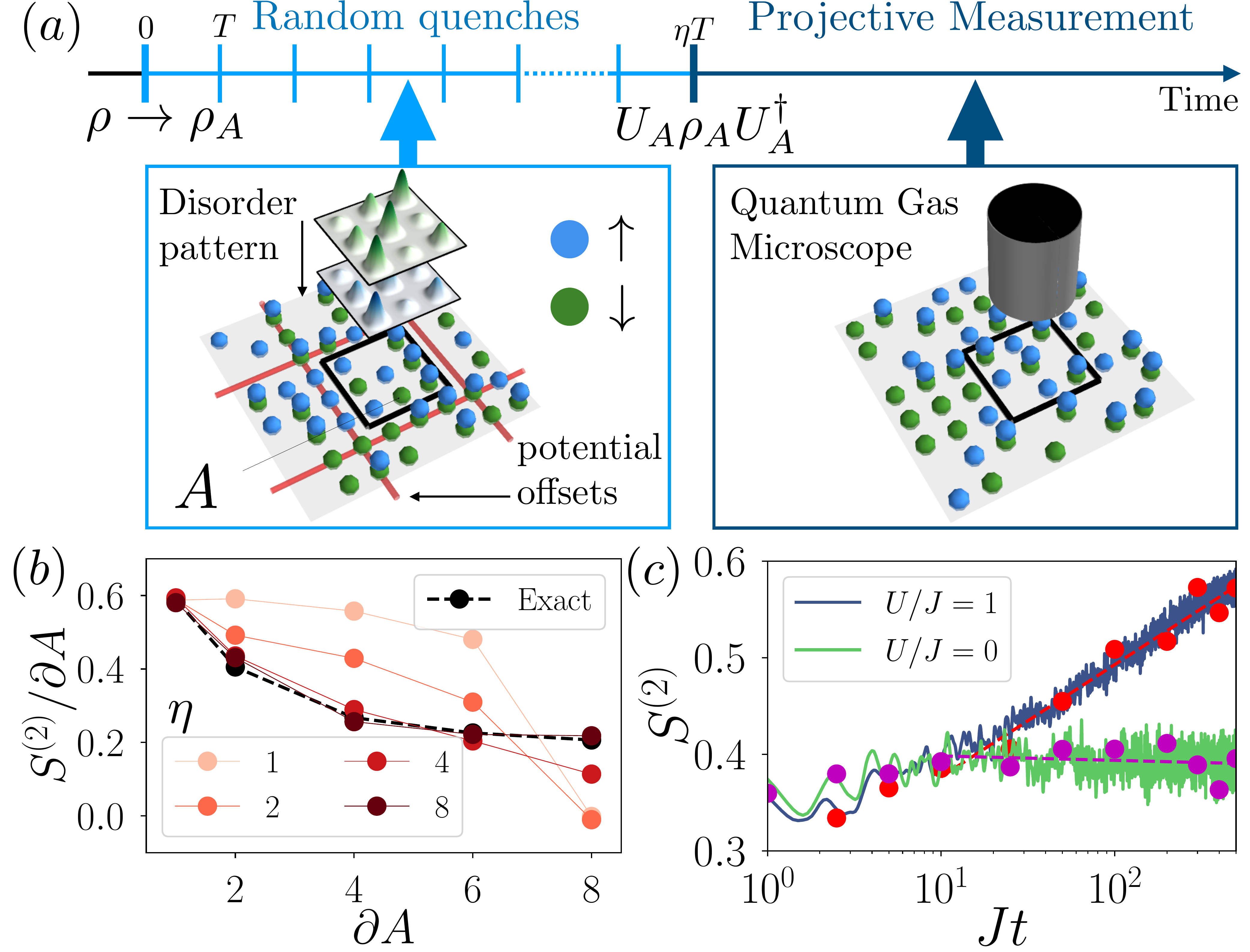} \caption{\textit{Measuring R{\'e}nyi entropies   via random quenches.}
(a) Experimental sequence: for a given reduced density matrix $\rho_{A}=\tr[\mathcal{S}\backslash A]{\rho}$  we apply (i) a random unitary $U_A$ realized by a series of $\eta$ random  quenches [c.f.~Eq.~\protect\eqref{eq:UA}], implemented using (spin-dependent) disorder potentials [c.f.~Eq.~\protect\eqref{eq:FH}]; this is followed by (ii) a projective measurement (read out) with a quantum gas microscope, to obtain $S^{(n)}(\rho_{A})$ from Eq.~\eqref{eq:purity}.
(b)  Within our protocol, we illustrate for the ground state of a 2D Heisenberg model  ($8\times8$ sites) area law scaling of $S^{(2)} \propto \partial A$ (with $\partial A$ the perimeter of area $A$), showing convergence with increasing $\eta$ to the exact value (black line). 
(c) For the many-body localized
phase of the 1D Bose Hubbard model ($10$
sites and $5$ particles), we illustrate a measurement of the logarithmic growth of $S^{(2)}(\rho_A)$ at half partition as a function of time.  The exact value of $S^{(2)}(\rho_A)$ 
(solid lines) is compared to the estimated values (dots).
The dashed lines are linear fits. The simulated experiments in (b-c) assume  $N_{U}=100$ random unitaries, and $N_{M}=100$ measurements per random unitary (see text).  \label{fig:setup}}
\end{figure}

Below we will describe a protocol for measuring R{\'e}nyi entropies $S^{(n)}(\rho_{A})$ based
on \emph{random measurements} realized as  \emph{random unitary operators}  applied to $\rho_A$ and subsequent measurements of a fixed observable~\cite{VanEnk2012}. In our approach the required
{random unitaries} are implemented using the same AMO toolbox which
underlies the preparation of quantum phases and dynamics (c.f.~Fig.\ \ref{fig:setup}). This enables
a physical implementation of the protocol, applicable to generic Hubbard
and spin models and in arbitrary dimension. We emphasize that in contrast
to recent protocols to measure $n$-th order R{\'e}nyi entropies, which
requires preparation of $n$ identical copies~\cite{Daley2012,Abanin2012,Islam2015,Kaufman2016},
a random measurement protocol requires only a single quantum system~\cite{VanEnk2012}, and thus can
be implemented directly with existing AMO and solid state platforms~\cite{Houck2012,Cai2013}. A central
aspect in any measurement scheme for R{\'e}nyi entropies, as for quantum state tomography~\cite{Gross2010,Cramer2010,Lanyon2016}, is scaling of the experimental effort with
size of the system of interest: below we provide a detailed analysis
and feasibility study of required resources in terms of number of
measurements and random unitaries, and verification of random unitaries
\cite{Vermersch2018}.

Random measurements to infer R{\'e}nyi entropies have been discussed in
	a quantum information context~\cite{VanEnk2012}. These consist in applying to $\rho_{A}$ a \emph{random unitary}
	matrix $U_{A}$ from the circular unitary ensemble (CUE) followed by a measurement in the fixed computational basis to access to the  outcome probabilities
	$\mathrm{P}(\vec{s})=\tr[]{ U_A \rho_A U_A ^\dagger \mathcal{P}_{\vec{s}}}$
	with  $\mathcal{P}_{\vec{s}} =\ketbra{\vec{s}}{\vec{s}}$  projectors onto the  basis states $\ket{\vec{s}}$.
	The extraction of the R{\'e}nyi entropies $S^{(n)}(\rho_{A})$ is then  based on the estimation of the statistical moments
	\begin{equation}
	\langle\mathrm{P}(\mathbf{s})^{n}\rangle= \langle\tr[]{ ( U_A \rho_A U_A ^\dagger )^{\otimes n} \mathcal{P}^{\otimes n}_{\vec{s}}} \rangle,  \label{eq:Pn}
	\end{equation}
	with $\langle\ldots\rangle$ the ensemble average over random unitaries.
	In order to obtain  $S^{(n)}(\rho_{A})$ from Eq.~\eqref{eq:Pn}, one relies on the statistical properties of the correlators between the matrix elements $u_{\vec{i}\vec{j}}$ of $U_A$. In particular, for  $n=2$, one exploits the identity
	\begin{equation}\label{eq:2design}
	\langle u_{\vec{si_{1}}}u_{\vec{si_{2}}}^{*}u_{\vec{si_{3}}}u_{\vec{si_{4}}}^{*}\rangle=\frac{\delta_{\vec{i_{1},i_{2}}}\delta_{\vec{i_{3},i_{4}}}+\delta_{\vec{i_{1},i_{4}}}\delta_{\vec{i_{2},i_{3}}}}{\hilbertn(\hilbertn+1)},
	\end{equation}
	with $\hilbertn$ the Hilbert space dimension of $A$, to obtain
	$\langle\mathrm{P}(\mathbf{s})^{2}\rangle = (1+\tr[]{\rho^2_A}) /(\hilbertn(\hilbertn+1))$~\cite{VanEnk2012}. Inverting this relation warrants  direct access to $S^{(2)}(\rho_{A})$ as a function of $\mathrm{P}(\mathbf{s})$~\footnote{By averaging estimated $S^{(2)}(\rho_A)$ obtained from different states $\bf{s}$, statistical errors are reduced (c.f.~below)}.
In the following, we  use that the required identities of  $n$-th order correlators of the CUE are reproduced by \emph{unitary $n$-designs}~\cite{Gross2007,Roy2009}, i.e.~ensembles of random unitary matrices approximating the CUE  by having the same correlators up to $n$-th order~\footnote{E.g.~ Eq.\eqref{eq:2design} is satisfied for $2$-designs.}. In contrast to the seminal  experiments measuring  $S^{(2)}(\rho_{A})$  in a BH model \cite{Islam2015} which rely on preparation of \emph{physical copies} of the quantum system \cite{Daley2012}, the present scheme works with single copies \cite{VanEnk2012}: The moments  \eqref{eq:Pn}  can be interpreted as  a replica trick to create $n$ \emph{virtual copies} [c.f. Eq.~\eqref{eq:Pn}]. We present additional details and a diagrammatic approach in the supplementary material (SM) \cite{SM}\nocite{Islam2015,Daley2012,Gray2017,Vermersch2018,Collins2010,DAlessio2013,DAlessio2014,Machado2017,Mezzadri2006}.

While in a quantum information context random unitaries  from unitary $n$-designs are generated as a sequence of random gates~\cite{Oliveira2007,Znidaric2008,VanEnk2012}, we show that such random unitaries can be realized with the existing AMO toolbox,  as a \textit{series of quenches} in interacting Hubbard and spin models with \textit{engineered disorder},
\begin{equation}
U_A= e^{-iH_A^\eta T} \cdots  e^{-iH_A^1T}, \label{eq:UA} 
\end{equation} 
followed by a readout with a quantum gas microscope (see Fig.~\ref{fig:setup}). Here, $H_A^j $ denotes the Hamiltonian for a given disorder pattern $j$.  In total, we consider $\eta$ quenches of duration $T$, with $T_{\rm tot}\equiv \eta T$ the total time. 
The questions to be addressed are: (i) the convergence to the CUE in terms of $n$-designs [c.f.~Eq.~(\ref{eq:2design})] with `depth'  $\eta$, in view of  experimentally available disorder Hamiltonians and experimental verification;  and (ii) the scaling of statistical errors with the number of applied random unitaries $N_{U}$ and the number of measurements per random unitary $N_{M}$. 
We emphasize the relation of (i) to the ongoing theoretical~\cite{DAlessio2013,DAlessio2014,Ponte2015,Gopalakrishnan2016,Machado2017}  and experimental~\cite{Bordia2017} investigation of thermalization dynamics of periodically driven quantum systems, and their connection to quantum chaos~\cite{Haake2010}. The type of problems, which can be addressed with our protocol are illustrated in Fig.~\ref{fig:setup}(b,c), with the simulation of the  \textit{measurement of an area law} for a 2D-Heisenberg model ~\cite{Song2011}, and  of the entropy growth in many-body localized~\cite{Basko2006,Bardarson2012,Serbyn2013,Nandkishore2015,Altman2015}  (MBL) dynamics in the Bose-Hubbard (BH) model, with details on the simulations presented below and  in the SM \cite{SM}.

{\it Protocol for the Fermi-Hubbard model} -- In view of recent progress in realizing the 2D Fermi Hubbard (FH) model~\cite{Haller2015,Greif2016,Boll2016,Cheuk2016}, we wish to illustrate the protocol for spinful fermions in a 2D optical lattices [c.f.~Fig.~\ref{fig:setup}(a)].  The FH Hamiltonian is
\begin{equation}
H_{F}= - t_F\sum_{\langle \vec{i,l}\rangle \in \mathcal{S},\sigma}c_{\vec{i}\sigma}^{\dagger}c_{\vec{l}\sigma}+U \sum_{\vec{i}\in \mathcal{S}}n_{\vec{i}\uparrow}n_{\vec{i}\downarrow}.\label{eq:FH}
\end{equation}
with hopping amplitude $t_F$, and interaction strength $U$. Here
 $c_{\vec{i},\sigma}^{(\dagger)}$ denote fermionic annihilation (creation)
operators at lattice site $\vec{i}=(i_x,i_y)$ and spin $\sigma\in\left\{ \uparrow,\downarrow\right\} $, and $n_{\vec{i}\sigma} = c_{\vec{i}\sigma}^{\dagger}c_{\vec{i}\sigma}$. We will add disorder below to realize $H_A^j$.

We assume that the (non-)equilibrium quantum many body state $\rho$ of interest has been prepared in 
the full system $\mathcal{S}$.
The experimental sequence to measure R{\'e}nyi entropies $S^{(n)}(\rho_A)$ of the reduced density matrix $\rho_{A}=\tr[\mathcal{S}\backslash A]{\rho}$   is  shown in Fig.~\ref{fig:setup}(a):
(i) Isolation of the partition $A$ of dimension ($L_x,L_y$)  and $L\equiv L_x L_y$ the number of isolated sites, is obtained via spatial addressing [c.f.~Fig.~\ref{fig:setup}(a)]. The Hamiltonian $H_{A}^{j}$ is realized as restriction $H_{A}^{j}=\left.H_{F}\right|_{A}+\sum_{\vec{i}\in A,\sigma}\delta_{\vec{i},\sigma}^{j}n_{\vec{i}\sigma}$  with random lattice offsets $\delta_{\vec{i},\sigma}^{j}$.
Due to particle and spin conservation in $H_F$, $U_A$ decomposes  into blocks with different
 particle number $N$ and magnetization $S_z$, $U_{A}=\bigoplus_{N,S_z}U_{A}^{(N,S_z)}$ and  $\rho_{A}=\bigoplus_{N,S_z}\rho_{A}^{(N,S_z)}$. Below we study in each block the realization of a random unitary $U_{A}^{(N,S_z)}$ from an $n$-design ($n=2,3,\ldots$) as function of $\eta$ and $T_{\rm tot}$.
(ii) Lattice site occupations $\vec{s}_{N,S_z}$ are measured  with a quantum gas microscope, where  $\vec{s}_{N,S_z}=(n_{\vec{i},\uparrow},n_{\vec{i},\downarrow})_\vec{i}$ determines $N=\sum_{\vec{i}\in A} (n_{\vec{i}\uparrow}+n_{\vec{i}\downarrow})$ and $S_z=\sum_{\vec{i}\in A}(n_{\vec{i}\uparrow}-n_{\vec{i}\downarrow})$. By repeating steps (i-ii) with the same  $U_A$, i.e.\ the same series of random quenches, to perform $N_M$ measurements, one estimates the probabilities $\probsU{\vec{s}_{N,S_z}}=\tr[]{U_A \rho_A U_A^\dagger \mathcal{P}_{\vec{s}_{N,S_z}}}$ with  $\mathcal{P}_{\vec{s}_{N,S_z}}$ the projector onto the Fock state $\ket{ \vec{s}_{N,S_z}}$ \footnote{Our protocol can be realized with other observables~\cite{Vermersch2018}}. Repeating this  for $N_U$ different unitaries, we estimate the ensemble averages $\probs[n]{\vec{s}_{N,S_z}}$, related to functionals of  $\rho_A$ \cite{VanEnk2012}.
Using  $1$- and $2$-design properties, we find
\begin{eqnarray}  
\probs{\vec{s}_{N,S_z}}&=& \frac{\tr{\rho^{(N,S_z)}_A}}{\hilbert} , \label{eq:trace} \\ 
\probs[2]{\vec{s}_{N,S_z}}&=& \frac{\tr{\rho^{(N,S_z)}_A}^2 + \tr{\rho^{(N,S_z)2}_A}}{\hilbert\left(\hilbert +1 \right)} \label{eq:purity},
\end{eqnarray}
 where $\hilbert$ is the Hilbert space dimension of the particle-spin block in the subsystem $A$. 
 Hence, from estimations of $\probs[n]{\vec{s}_{N,S_z}}$ ($n=1,2$),   $\tr{\rho^{(N,S_z)2}_A}$\ can be extracted. By summation over all blocks, one obtains  the total purity $p_2\equiv\tr{\rho^{2}_A}=\sum_{N,S_z}\tr{\rho^{(N,S_z)2}_A}$ and finally $S^{(2)}(\rho_A)$. 
Higher order ensemble averages $\probs[n]{\vec{s}_{N,S_z}}$ are related to higher order powers $\mathrm{Tr}(\rho_A^{(N,S_z)n})$ \cite{Vermersch2018}.

\textit{Generation of random unitaries} --
Below we  present a numerical study of  generation of approximate unitary $2$-designs~\cite{Dankert2009,Ohliger2013,Brandao2016,Banchi2017}, focusing on convergence of the $U_A$ [c.f.~ Eq.~(\ref{eq:UA})] to the CUE as function of time $T_{\rm tot}=\eta T$, and depth $\eta$ of the `random circuit'.  
While the full system $\mathcal{S}$ can be arbitrary large, we emphasize that --- in view of the scaling of statistical errors with the partition size $A$ (see below) --- the applicability of the protocol in an actual experiment will \textit{a priori} be limited to domains $A$ of moderate size, which can be simulated numerically.
 Here, we present results for the Heisenberg model in 1D and 2D, which allows larger partition sizes, and we refer to Ref.~\cite{Vermersch2018} for the FH model. 
 The Hamiltonian is 
  $H_{h}= J \sum_{\langle \vec{i}\vec{l}\rangle} \vec{\sigma}_\vec{i}.\mathbf{\sigma}_\vec{l}$, as obtained from Eq.~(\ref{eq:FH}) in the limit $U\gg t_F$ at half filling (alternatively  with Rydberg atoms~\cite{Gorshkov2011} or trapped ions~\cite{Porras2004}).  Here, $\sigma_\mathbf{i}$ are the Pauli matrices, and $J=t_F^2/U$.
  To realize random quenches, we consider disorder potentials  $\delta_\vec{i}^j=\delta^j_{\vec{i}\uparrow}-\delta^j_{\vec{i}\downarrow}$ drawn for each quench $j$ from a normal distribution with standard deviation $\delta$, i.e.~$H_{A}^{j}\equiv \left.H_{h}\right|_{A}+\sum_{\vec{i}\in A}\delta_\vec{i}^{j}\sigma_\vec{i}^z$ \footnote{We have found that random unitaries can be realized using Aubry-Andr\'{e} potentials \cite{Aubry1980} with similar convergence times.}. 
 \begin{figure}[t]
	\includegraphics[width=0.97\columnwidth]{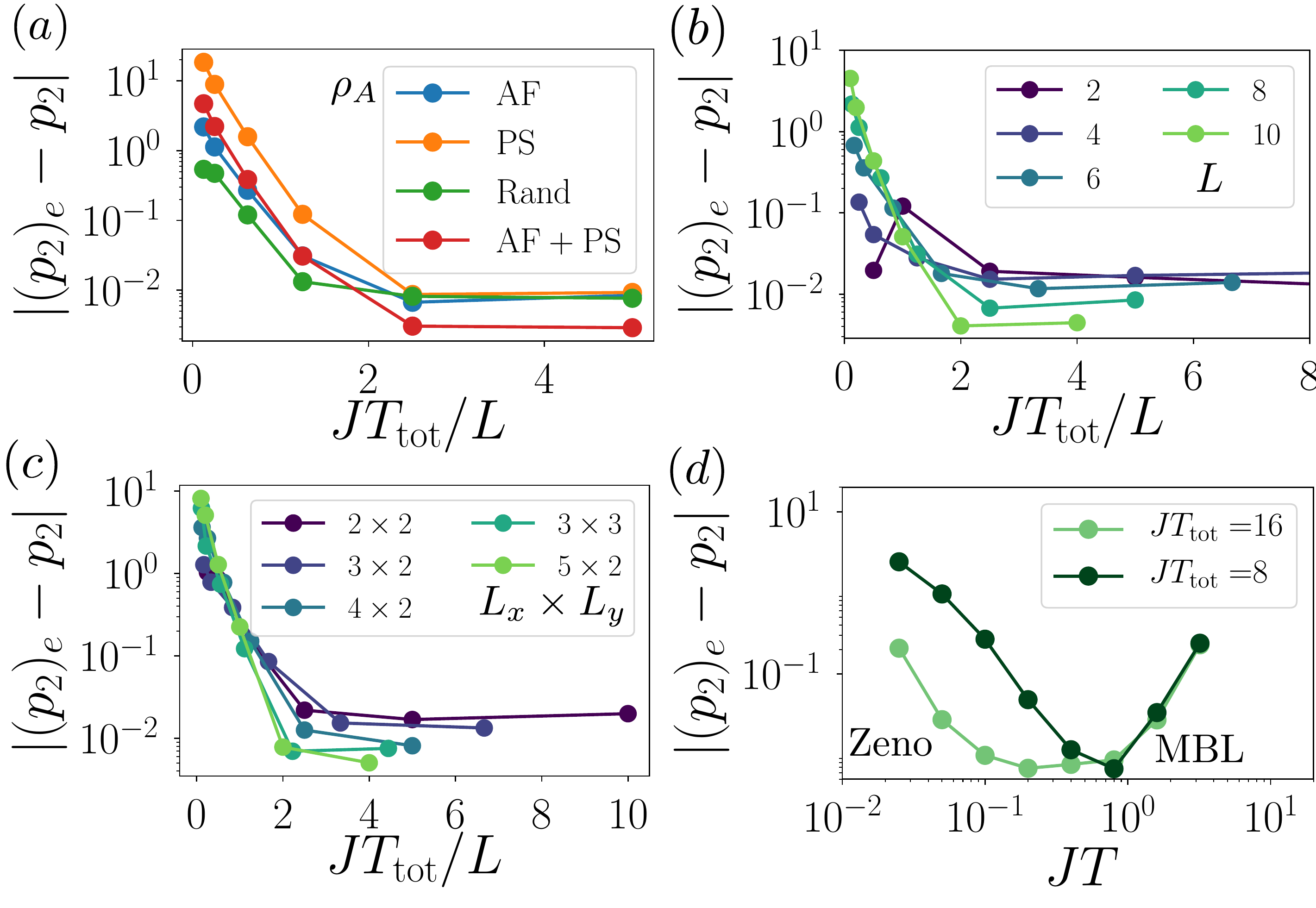}
	\caption{\textit{Creation of approximate  $2$-designs in  the  Heisenberg model}. 
		(a) Average error of the estimated purity $|(p_2)_e - (p_2)|$ for a uni-dimensional partition of size $L=8$ and various test states: an antiferromagnetic state $\ket{\psi_\mathrm{AF}}$, the phase separated state $\ket{\psi_\mathrm{PS}}=\prod_{\vec{i},i_x\le L_x/2}\ket{\downarrow}_\vec{i}\prod_{\vec{i},i_x> L_x/2}\ket{\uparrow}_\vec{i}$, a pure  random state $\ket{\psi_\mathrm{rand}}$ with $S_z=0$, and the mixed state $\rho_A=\frac{1}{2}(\ket{\psi_\mathrm{AF}}\bra{\psi_\mathrm{AF}}+\ket{\psi_\mathrm{PS}}\bra{\psi_\mathrm{PS}})$.
	(b-c) Error for  $\rho_A=\ket{\psi_\mathrm{AF}}\bra{\psi_\mathrm{AF}}$  for (b) uni-dimensional partitions ($L=L_x$) and (c) two-dimensional partitions ($L=L_xL_y$).
	(d) Optimization of the quench time $JT$ for fixed total time $T_\mathrm{tot}$ and disorder strength $\delta =J$. 
	For all panels, we average over $N_U=500$ unitaries and consider $N_M=\infty$.\label{fig:H}}
\end{figure}

Fig.~\ref{fig:H}(a-d)  shows the error of the estimated purity $(p_2)_e$ of various test states $\rho_A$ (defined in the caption) for partitions $A$ of various sizes $L$ in 1D ($L=L_x$) and 2D ($L=L_xL_y$)~\footnote{Note that $(p_2)_e$ can be larger than $1$ when $T_\mathrm{tot}\to0$}. According to panels~(a,b,c), for a fixed quench time $JT=1 $ and disorder strength $\delta =  J$, the  error decreases exponentially with growing $J T_\mathrm{tot}/L=\eta/L$ towards a plateau, which corresponds to the statistical error threshold (see below).
Thus our results indicate `efficient' convergence of $U_A$ to an approximate $2$-design, after a total time $T_{\textrm{tot}}$ which scales linearly with $L$, as in conventional random circuits based on engineered gates~\cite{Dankert2009,Ohliger2013,Brandao2016}.  Note that our simulations show that product states, which are prepared in an experiment with high fidelity, provide  good indicators of convergence of the generated unitaries. 

For a given total time $T_\mathrm{tot}$, set in a experiment by the finite coherence time, we show in panel Fig.~\ref{fig:H}(d) the existence of an optimal quench time $ J T\approx 1$ to minimize errors.
This reflects the trade-off between the requirements of (i) to evolve the system for each quench $j$ during a time sufficiently large compared to   timescales $J^{-1},\delta^{-1}$ set by the Hamiltonian~\cite{Facchi2005}, i.e.~to prevent a quantum Zeno effect, and (ii) to change the disorder pattern frequently to prevent 
localization. It also exists an optimal disorder strength $\delta\approx J$ \cite{SM}, resulting from a tradeoff  between localizing effects in the limit $\delta\gg  J$ and a vanishing random component of the applied quenches in the limit $\delta \ll J$.
We note  that the use of a single disorder pattern, combined with random quench times $T\to T_j$, represents another possibility  to generate the required random unitaries~\cite{SM}.

Our findings, in particular the convergence to approximate $2$-designs and the corresponding scalings, also apply to generic Fermi and Bose Hubbard models, and  quantum Ising models~\cite{Vermersch2018}.  
Moreover, we emphasize that (i) our measurement scheme does not rely on the knowledge of the applied unitaries $U_A$ and (ii) -- with respect to state-of-the-art AMO setups --  the measurement protocol is robust against imperfect reproducibility of the generated unitaries, finite detection fidelity and decoherence~\cite{Vermersch2018}.
While we are interested in this work in the limit of large times $T_\mathrm{tot}$ where approximate $2$-designs are created (as part of our measurement scheme), we finally remark that random quenches in AMO systems provide a platform to study fast thermalization dynamics towards quantum chaos~\cite{Vermersch2018} and the entanglement growth, associated with random time evolution~\cite{Nahum2017}.

\textit{Statistical errors} --
We now discuss the statistical errors due to a finite number of random unitaries $N_U$ and of measurements $N_M$ per unitary.
\begin{figure}[t]
	\includegraphics[width=0.97\columnwidth]{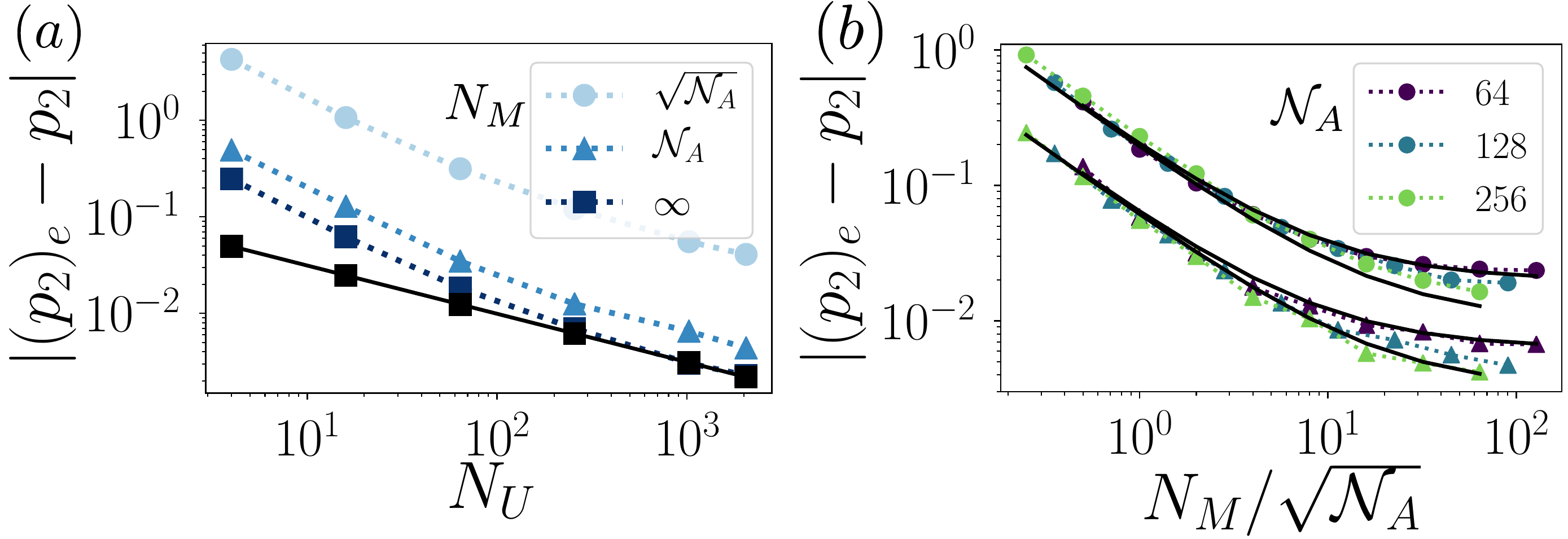}
	\caption{\textit{Scaling of statistical errors}. (a) Average statistical error of the estimated purity as a function of $N_{U}$
		for various $N_{M}$, $\hilbertn=256$. (b)
		Error as a function of $N_{M}$, for different $\mathcal{N}_A$, showing birthday paradox scaling $N_M/\sqrt{\hilbertn}$. Circles represent $N_{U}=100$ and triangles $N_{U}=1000$.
		The unitaries are sampled from the CUE numerically~\cite{Mezzadri2006}. The black lines represent the expressions given in the text and Ref.~\cite{Vermersch2018}.}
	\label{fig:errors} 
\end{figure}
For simplicity, we assume that 
$\rho_{A}=\rho_{A}^{(N,S_z)}$ describes a state
in a single spin-particle sector with dimension $\hilbert$, where
random unitaries from the CUE are created.
Since the following discussion is not specific to an underlying model, we also drop the labels $(N,S_z)$.
In Fig.~\ref{fig:errors}(a), the average error of
the purity is shown as a function of $N_{U}$, decreasing as $1/\sqrt{N_{U}}$
for fixed $N_{M}$. In panel (b), it is represented as a function of $N_M$, for $N_U=100$ and $1000$.
We find that for $N_U \gg 1$, the error scales as $|({p_2})_{e}-p_2|\sim\left(C_2+{\mathcal{N}_A}/{N_{M}}\right)/{\sqrt{\mathcal{N}_AN_{U}}}$, where $C_2=\mathcal{O}(1)$ is largest for pure states. 
The results are confirmed by the analytical study presented in Ref.~\cite{Vermersch2018}. The first term, independent of $N_{M}$,
arises from the finite value of $N_U$ ~\cite{VanEnk2012}.
The second originates from the finite number $N_{M}$ of measurements. It leads to a requirement of $N_{M}\sim\sqrt{\mathcal{N}_A}$ to determine the purity up to an error of the order $1/\sqrt{N_{U}}$. This scaling is directly related to the statistics of doublons obtained when
sampling a discrete variable (the birthday paradox \cite{Blinder2013}).

The total number of measurements $N_MN_U$ scales polynomially with the Hilbert space dimension $\hilbertn$, and thus exponentially with the size of $A$ (independently of the total system $\mathcal{S}$).
However, compared to quantum state tomography, the exponent
is favorable and allows to perform measurements of $S^{(2)}(\rho_A)$ for subsystem sizes, which are for instance compatible with the examples in Fig.~\ref{fig:setup}.

\textit{Application to physical examples} --
We conclude our discussion by presenting applications of the protocol investigating  entanglement properties of quantum many-body states $\ket{\psi}$.
As  first example, we demonstrate in Fig.~\ref{fig:setup}(b) the \textit{measurement of an area law} in a 2D Heisenberg model. 
We consider a system $\mathcal{S}$ prepared in the $S_z=0$ ground state $\ket{\psi}$ of $H_h$ on an $8\times 8 $ square lattice, obtained numerically with DMRG \footnote{With maximal bond dimension $4000$ and $17$ sweeps.}. For rectangular partitions $A$ with size $L=L_x L_y$ placed at the center of the system, we estimate  the second R{\'e}nyi entropy $S^{(2)}(\rho_A)$ of the reduced density matrices $\rho_A=\tr[\mathcal{S}\backslash A]{ \ket{\psi}\bra{\psi}}$ as a function  of the partition boundary $\partial A=2(L_x+L_y-2)$. We observe that the estimated R{\'e}nyi entropy converge to the area law result~\cite{Song2011} with increasing number of quenches $\eta$. The quench parameters are $\delta=J=1/T$. Note that we have used here a finite number of unitaries $N_U=100$, and a finite number of measurements $N_M=100$.  
As  second example, Fig.~\ref{fig:setup}(c) shows for a $1$D Bose Hubbard model the \textit{entanglement growth in the many-body localized (MBL) phase}~\cite{Bardarson2012,Serbyn2013}, with details on the model and parameters summarized in the SM~\cite{SM}.
According to Fig.~\ref{fig:setup}(c), the estimated second order R{\'e}nyi entropy as a function of time clearly allows to distinguish MBL from Anderson localization.

\begin{figure}
	\includegraphics[width=0.93\columnwidth]{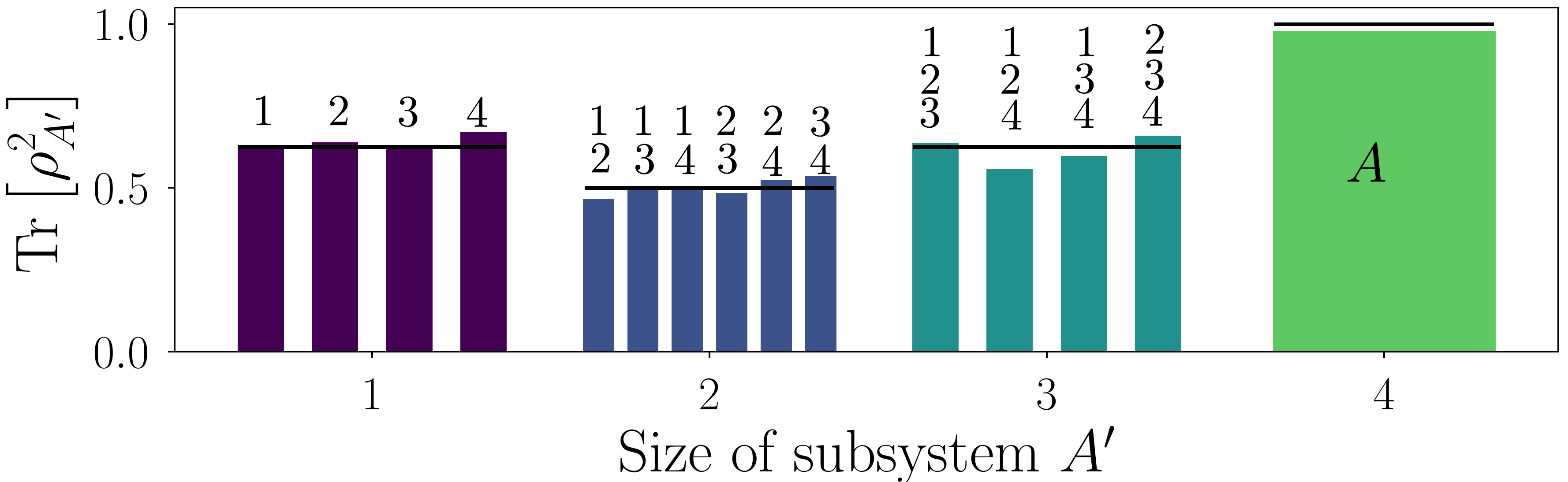}
	\caption{\textit{Protocol with local unitaries.} Purity of all (sub-) systems $A'\subseteq A$ with $N_U =2N_M=100$. 
		The numbers refer to the indices $i=1,..,L$ contained in $A'$, the green bar to $A'=A$. The black lines indicate the exact values.
	}
	\label{fig:local} 
\end{figure}

{\it Protocol based on  local unitaries} -- The measurement scheme described above relies on global  entangling unitaries acting on the entire Hilbert space. As an alternative, we can  use \textit{local unitaries}, which act  individually on local constituents $i=1,\dots,L$ (e.g.~spins)  of $A$.  Here, the unitary $U_A$ is given as a product $U_A = u_1\otimes ... \otimes u_L$ where  each $u_i$ is independently drawn from a unitary 2-design in the local Hilbert space of dimension $d$. In the case of a spin system,  the $u_i$ can be viewed as random single spin rotation on the Bloch sphere. As above, from  measurements of the local spin configuration
with outcome $\vec{s}=(s_i)_{i=1,..,L}$,  we compute the statistical moments $\langle \text{P}(\vec{s})^n \rangle$.
We find $\langle \text{P}(\vec{s}) \rangle = 1/d^L$~\footnote{The unitaries $U_A$ form  a $1$-design.} and, using the $2$-design properties of $u_i$, 
\begin{eqnarray}
\langle \text{P}(\vec{s})^2 \rangle = \frac{\sum_{A' \subseteq A}\mathrm{Tr}(\rho_{A'}^2)}{d^L(d+1)^L} \, .\label{eq:localp}
\end{eqnarray}
Here, we sum over all subsystems $A' \subseteq A$, including the empty subsystem with $\tr{\rho_\emptyset^2}\equiv 1$.
Since the unitaries act only locally,  Eq.~\eqref{eq:localp} holds for each subsystem $A'$. This allows to  reconstruct recursively all  purities  $\mathrm{Tr}(\rho_{A'}^2)$ for $A' \subseteq A$. Local unitaries allow thus to infer more information from the measurement than global unitaries. This is illustrated in Fig.~\ref{fig:local} for $L=4$ spins initialized in the $W$-state.  We note however, that due to the recursive reconstruction of the purities from Eq.~\eqref{eq:localp}, this protocol is more prone to statistical errors~\cite{SM}.

{\it Conclusion and Outlook} -- 
Our protocol allows the measurement of R{\'e}nyi entropies based on single copies in existing AMO setups:  for example, to obtain the purity of $\rho_A$ of a partition $A$ with $L=14$ spins,  as part of an arbitrarily large many-body system, one needs for an  accuracy of $\sim5\%$ to create unitaries during a time $JT_\mathrm{tot}\sim 25$, and to perform $N_M=500$ measurements for $N_U=100$ unitaries. 
While we have focused on measurement of second order R{\'e}nyi entropies, higher order entropies are also accessible although with increasing statistical errors \cite{Vermersch2018}, which provides an interesting perspective to extend the protocol to von Neumann entropies, or the entanglement spectrum~\cite{Li2008,Pichler2016,Dalmonte2017}. 

\begin{acknowledgments}
We thank the M.~Lukin,  M.~Greiner, M.~Hafezi group members, and J.~Eisert, C.~Roos, P.~Jurcevic, G.~Pagano, W.~Lechner, M.~Baranov, H.~Pichler, P.~Hauke, M.~\L\k{a}cki, and D. ~Hangleiter for  discussions.
The DMRG and exact diagonalization simulations were performed using the ITensor library (http://itensor.org) and QuTiP~\cite{Johansson20131234}, respectively.
Work in Innsbruck is supported by the ERC Synergy Grant UQUAM and the SFB FoQuS (FWF Project No. F4016-N23). JIC acknowledges support from the ERC grant QUENOCOBA.
\end{acknowledgments}

\input{20171117PRL_MT.bbl}

\appendix
 \section{Diagrammatic approach on virtual copies}
 
 In this section, we show how to relate the values of $ \langle\mathrm{P}(\mathbf{s})^{2}\rangle= \langle\tr[]{ ( U_A \rho_A U_A ^\dagger )^{\otimes n} \mathcal{P}^{\otimes 2}_{\vec{s}}} \rangle$ to functionals of $\rho_A$, based on a diagrammatic approach involving  `virtual' copies of $\rho_A$.   
 We assume that $\rho_A$ is defined in a Hilbert space $\mathcal{H}$ with dimension $\hilbertn$ and basis $\left\{ \ket{\vec{s}}  \right\}$ and that the random unitaries $U_A$ are drawn from a unitary $2$-design, such that Eq.~(2) of the main text (MT) holds. The projectors $\mathcal{P}_{\vec{s}} =\ketbra{\vec{s}}{\vec{s}}$ describe direct measurements of occupations of basis states. 
 
\begin{figure}[h!]
 	\includegraphics[width=1\columnwidth]{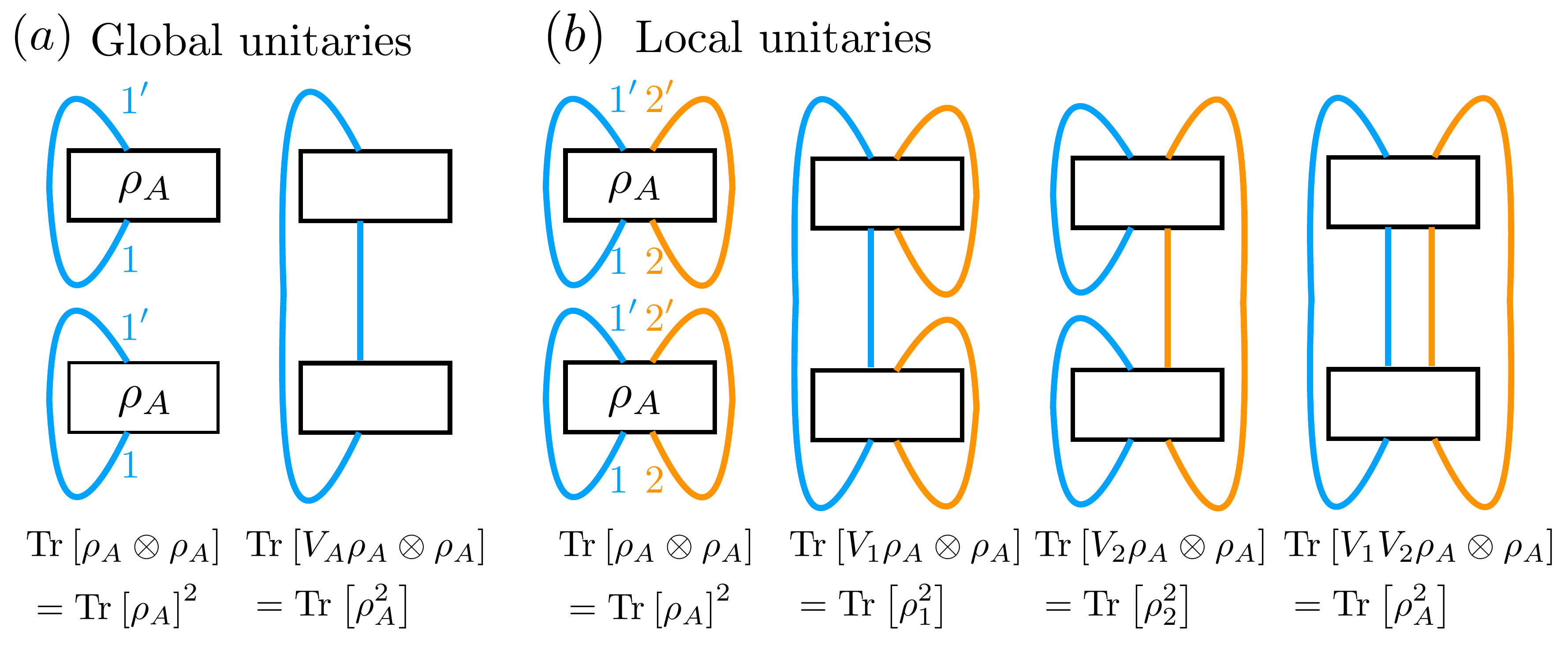}
 	\caption{{\it Random measurements on virtual copies}. The ensemble average $ \langle\mathrm{P}(\mathbf{s})^{2}\rangle$ can be evaluated using a simple diagrammatic approach involving two virtual copies of $\rho_A$. In panel a), we consider the case of global random unitaries $U_A$, in panel b) $L=2$ local random unitaries, i.e.\ $U_A=U_1 \otimes U_2$. Here,  $\rho_1=\tr[2]{\rho_A}$ ($\rho_2=\tr[1]{\rho_A}$)  denotes the reduced density matrix of the first (second) constituent. \label{fig:SM_copies}}
 \end{figure}

 We note that the measurement of the second order R\'enyi entropy in Ref.~\cite{Islam2015} is based on the physical realization of a swap operator $V_A$ on two `real' copies of $\rho_A$ via a beam splitter operation \cite{Daley2012}. Here, we show that  the ensemble average $ \langle\mathrm{P}(\mathbf{s})^{2}\rangle$ can be understood as an expectation value of $V_A$ applied to two `virtual' copies of $\rho_A$.  
 Similar to Ref.~\cite{Daley2012}, we define $V_A$ on the product space $\mathcal{H} \otimes \mathcal{H}$ by 
 \begin{align}
 V_A\ket{\vec{s}}_1\otimes \ket{\vec{t}}_2 \equiv \ket{\vec{t}}_1\otimes \ket{\vec{s}} _2
 \end{align} 
such that $\bra{\vec{s}'}\otimes \bra{\vec{t}'} V_A\ket{\vec{s}}\otimes \ket{\vec{t}}= \delta_{\vec{s}',\vec{t}} \delta_{\vec{t}',\vec{s}}$. By comparison with 
Eq.~(2) of the MT we thus find that 
  \begin{align}
 \langle \text{P}(\vec{s})^2 \rangle
  &=\frac{\tr[]{ (\mathbb{1}+V_A ) \rho_A\otimes\rho_A} }{\hilbertn(\hilbertn+1)}\nonumber \\
  &= \frac{\tr{\rho_A}^2+\tr{\rho^2_A}}{\hilbertn(\hilbertn+1)} \;, \label{eq:psqr}
 \end{align}
 where $\mathbb{1}$ is the identity operator, with  $\bra{\vec{s}'}\otimes \bra{\vec{t}'} \mathbb{1}\ket{\vec{s}}\otimes \ket{\vec{t}}= \delta_{\vec{s}',\vec{s}} \delta_{\vec{t}',\vec{t}}$.  Eq.~\eqref{eq:psqr} can be visualized using a simple diagrammatic approach (see Fig.~\ref{fig:SM_copies} and also Ref.~\cite{Gray2017}): To evaluate $\langle\mathrm{P}(\mathbf{s})^{2}\rangle$, we  draw the two virtual copies of $\rho_A=\sum_{\vec{s},\vec{s}'}   (\rho_A )_{(\vec{s},\vec{s}')}\ketbra{\vec{s}}{\vec{s}'}  $,  as boxes, each with two legs, corresponding to the primed and unprimed indices, respectively. Then we connect  unprimed and primed legs in all possible ways, to  contract the indices. This results in the two diagrams presented in Fig.~\ref{fig:SM_copies} (a) which correspond to the two summands  in Eq.~\eqref{eq:psqr}.
 
Now, we consider the variant of the protocol presented in the MT which is based on local random unitaries. Here, the random unitaries take the form $U_A=\bigotimes_{l=1}^L U_l$ with $U_l$ ($l=1,\dots,L$) drawn independently from unitary $2$-designs defined on the Hilbert spaces $\mathcal{H}_l$ of the local constituents with dimension $\hilbertl$  \footnote{Note that locality is defined in terms of the random unitaries $U_A=\bigotimes_{l=1}^L U_l$. Each factor $U_l$ defines a local constituent.}.   We assume that $\left\{ \ket{\vec{s}}  \right\}=\left\{ \ket{\vec{s}_1,\dots,\vec{s}_L}  \right\}$ denotes the product basis in $\mathcal{H}=\bigotimes_{l=1}^L \mathcal{H}_l$ and define the restricted swap operator 
 \begin{align}
V_l&\ket{\vec{s}_1,\dots,\vec{s}_l,\dots,\vec{s}_L}_1\otimes \ket{\vec{t}_1,\dots,\vec{t}_l,\dots,\vec{t}_L}_2 
\nonumber \\ & \equiv   \ket{\vec{s}_1,\dots,\vec{t}_l,\dots,\vec{s}_L}_1\otimes \ket{\vec{t}_1,\dots,\vec{s}_l,\dots,\vec{t}_L}_2,
\end{align} 
swapping only indices of the $l$-th constituent. Using the $2$-design properties (Eq.~(2) of the MT) of the $U_l$ ($l=1,\dots,L$), we find similar as in the global case 
   \begin{align}
 \langle \text{P}(\vec{s})^2 \rangle
 &=\frac{\tr[]{ \prod_{l=1}^{L}  (\mathbb{1}+V_l ) \; \rho_A\otimes\rho_A} }{\prod_{l=1}^{L} \hilbertl(\hilbertl+1)}  
 \label{eq:psqrloc}
 \end{align} 
 which reduces  to Eq.~(7) of the MT.
To visualize this in the diagramatic language developed above, we draw now for each virtual copy of $\rho_A$ boxes with $2L$ legs, corresponding to the $L$ primed and $L$ unprimed indices of $\rho_A$. Then, we  connect, for each local constituent separately, primed and unprimed legs to contract indices. For the case $L=2$, the resulting diagrams are shown in  Fig.~\ref{fig:SM_copies} (b) and correspond to the four summands in the nominator of Eq.~\eqref{eq:psqrloc}.

Finally, we note that the  diagramatic approach can be extended to $n>2$ to evaluate $ \langle\mathrm{P}(\mathbf{s})^{n}\rangle$ in local and global case. Furthermore, to evaluate ensemble averages of outcome probablities of random measurements of arbitary observables $\mathcal{O}$ \cite{Vermersch2018} (described by projectors $\mathcal{P}_{\mathcal{O}}$ with $\tr[]{\mathcal{P}_{\mathcal{O}}}>1$) 
a comprehensive graphical calculus for arbitrary moments of the CUE, developed in Ref.~\cite{Collins2010}, can be used.

\section{Random unitaries from random quenches in 1D and 2D Heisenberg models}

In this section, we complement the study of the convergence to $2$-designs for the Heisenberg model, as presented in the main text. 
We first discuss the optimization of the disorder strength $\delta$. We then present the possibility to create $2$-designs using a single disorder pattern.

\begin{figure}[h!]
	\includegraphics[width=1\columnwidth]{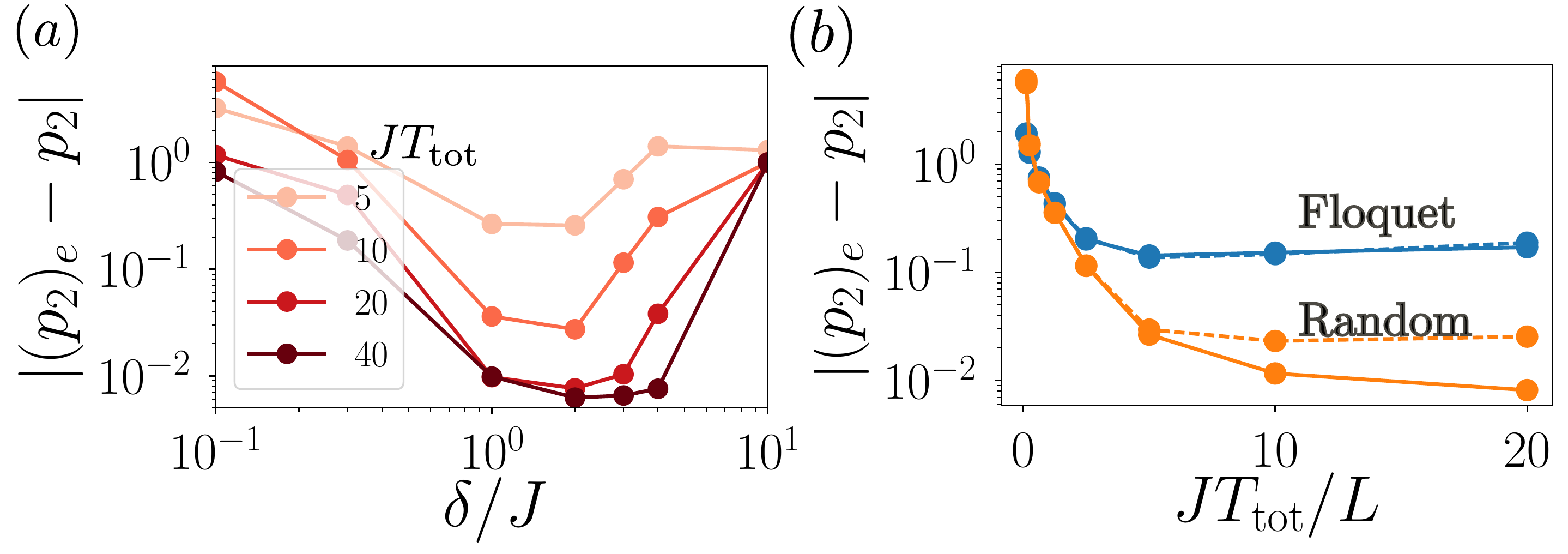}
	\caption{{\it Convergence to a unitary $2-$design in the 1D Heisenberg model}. We consider an antiferromagnetic state, $L=8$, $N_U=500$, and $N_M=\infty$.
		(a) Influence of the disorder strength $\delta$ for different $T_\mathrm{tot}$, showing an optimum at $\delta\approx J$.
		(b) Convergence to the CUE with a single disorder pattern $\delta_j$ and random times $T_j$. The dashed lines show the error for $N_U=100$.\label{fig:SM_H}}
\end{figure}

The optimization of random quenches with respect to $\delta$  is shown in Fig.~\ref{fig:SM_H} (a) for the antiferromagnetic state $\ket{\psi_\mathrm{AF}}$, $(L_x,L_y)=(8,1)$, and different times $T_\mathrm{tot}=\eta/J$. The error of the estimated purity is minimal around $\delta\approx J$.
Overall, we remark that the convergence to CUE is favored when all  relevant frequencies associated with the quenches are of the same order of magnitude (here $ J\approx \delta \approx 1/ T$).

For simplicity, we present in the MT the case where for each quench $j$, the applied disorder pattern $\delta^j_\vec{i}$ is not correlated with the previous realizations $j'<j$. This requires that the source of disorder (as implemented for instance with spatial light modulators (SLM) or speckle patterns in AMO systems) is dynamically reconfigurable.
In Fig.~\ref{fig:SM_H}(b), we show that random unitaries converging to $2$-designs can be also realized using a single disorder pattern $\delta_i$, which is drawn from a normal distribution of standard deviation $\delta=J$, and applied every second quench: $\delta_\vec{i}^j=\delta_\vec{i} \mathrm{mod}(j,2)$, provided the quench times $T\to T_j$ depend on $j$ and are random (here drawn for a uniform distribution in the interval $[0,2J^{-1}]$). 

Note that in the case of a constant quench time $JT_j=1$ (blue lines), corresponding to a Floquet system of period $2T$, the error remains large ($\sim 10^{-1}$) and does not depend on the number of unitaries $N_U=100,500$, i.e is not due to statistical errors. 
We attribute this to the slow thermalization dynamics of Floquet systems~\cite{DAlessio2013,DAlessio2014,Machado2017}, occurring at $JT_\mathrm{tot}\gg 1$ (which is not visible in Fig.~\ref{fig:SM_H}).

\section{Details on the Bose-Hubbard simulation}

In this section, we give additional details on Fig.~1(c) of the MT, displaying a the simulation of the measurement of the  entanglement growth in the MBL phase.
The BH Hamiltonian governing the dynamics is given by
\begin{align}
H_{B}=&-J\sum_{i\in \mathcal{S}}\left(a_{i+1}^{\dagger}a_{i}+\text{h.c.}\right)+\frac{U}{2}\sum_{i\in \mathcal{S}}n_{i}(n_{i}-1) \nonumber \\ & + \sum_{i\in \mathcal{S}} \delta_i n_i  \nonumber
\end{align}
with hopping $J$, onsite interaction $U$ and local disorder potentials $\delta_i$. Here,  $a_i$ ($a_i^\dagger$) denote bosonic annihilation (creation) operators and $n_i=a_i^\dagger a_i$ the local number operators. 
We consider a system with  $L_\mathcal{S}=10$ sites and $N_\mathcal{S}=5$ particles. We calculate its time evolution, via a Matrix-Product-State (MPS) simulation (truncation error $10^{-10}$, time step $0.1/J$), 
for  static disorder potentials $\delta_{i}$ uniformly distributed in $[-10J,10J]$  in the Anderson-localized
($U/J=0$) or many-body localized phase ($U/J=1$).
 We then obtain the second order R{\'e}nyi entropy $S^{(2)}(\rho_A)$, at half partition $A$, as a function of time $t$, and averaged over $250$ disorder realizations (solid lines).

To simulate the measurement scheme, we apply to $\rho_A$, which is extracted from the MPS simulation at certain times $t$, a series  ($j=1,..,\eta)$, with $\eta=20$, of random quenches governed by $H_{A}^{j}=-J\sum_{i \in A}\left(a_{i+1}^{\dagger}a_{i}+\text{h.c.}\right)+{U}/{2}\sum_{i \in A}n_{i}(n_{i}-1)+\sum_{i\in A}\delta_{i }^{j}n_{i}$ with (weak) disorder patterns $\delta_{i}^{j} $ drawn for each quench from a normal distribution with standard deviation $\delta=J$.
 The interaction during the random quenches is chosen to be $U=J$.
 The corresponding estimated R{\'e}nyi entropies, represented as circles, clearly enable to distinguish between Anderson- and many-body localized regime. 
 The convergence properties of random unitaries generated in the BH model are discussed in the companion paper~\cite{Vermersch2018}.

 \section{Statistical errors using local random unitaries}
 
  \begin{figure}[t!]
	\includegraphics[width=0.95\columnwidth]{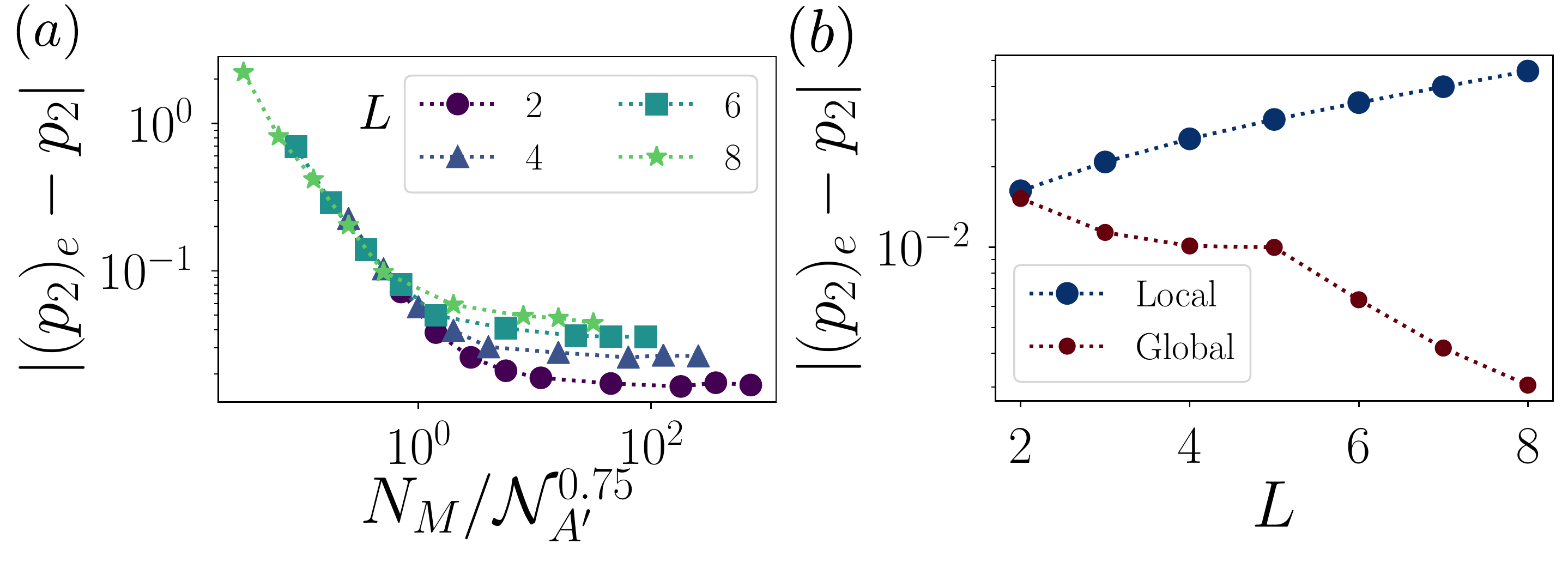}	
	\caption{\textit{Scaling of statistical errors in the local protocol.}  We consider 
		a spin-$1/2$-chain with $L$ spins and total Hilbert space dimension $\mathcal{N_A}=2^L$. (a) Error as a function of $N_{M}$, for various $L$, exhibiting the scaling $\hilbertn^{0.75}/N_M$  for $N_M\ll \mathcal{N}_A$. b) Comparison of the error in local and global protocol in the limit $N_M\rightarrow \infty$ as a function of the number of spins $L$.  $N_{U}=1000$ unitaries were drawn directly from the CUE \cite{Mezzadri2006}.}
	\label{fig:errors_local} 
\end{figure}

 We discuss now statistical errors involved in the estimation of the purity in the protocol based on unitaries $U_A=\bigotimes_{i} u_i$ with $u_i \in \text{CUE}(d_i)$ acting on a local constituent $i$ of the subsystem $A$ with  local Hilbert space dimension $d$.  As an example, we consider  a spin-$1/2$-chain with $L$ spins ($d_i=2$)  and total Hilbert space dimension $\mathcal{N}_A=2^L$. Note that the numerical analysis of statistical errors in the protocol based on global unitaries in the MT is complemented and extended by an analytical treatment  in Ref.~\cite{Vermersch2018}.

 In Fig.~\ref{fig:errors_local} (a), we display the average statistical error of the estimated purity of a reduced density matrix $\rho_A$ as a function of the number of measurements $N_M$ per random unitary, for various subsystem sizes and a fixed number of random unitaries $N_U=1000$. 
 In the limit $N_M \ll \mathcal{N}_A$ we find  numerically   a scaling of the statistical error of the estimated purity $ \sqrt{N_U} |(p_2)_e -p_2| \sim \mathcal{N}_A^\kappa/N_M $ with $\kappa =0.75 \pm 0.1$.  Compared to the global protocol (scaling exponent $\kappa = 1/2$) the error is hence increased. In Fig.~\ref{fig:errors_local} (b), we further observe  that in the limit   $N_M \rightarrow \infty$, the error grows with increasing system size $L$, which is contrary to the global protocol (see also Fig.~3 MT).
 Both results are  explained by the  fact that in the local protocol  the purity of the reduced density matrix $\rho_A$ of a subsystem $A$ is   recursively determined from the purities of the reduced density matrices $\rho_{A^\prime}$ of all subsystems $A^\prime \subset A$.  Hence, their statistical errors add  up. For larger systems, the number of involved subsystems increases, causing the growing statistical error. 
 
 To summarize, we find that the {\em local protocol}, compared to the global one,  is more prone to statistical errors and requires thus more measurements per random unitary to obtain the purity of the reduced density matrix $\rho_A$ of a subsystem $A$ up to a given error.  However,  we obtain in addition the purities of all reduced density matrices $\rho_{A'}$ of subsystems $A'\subseteq A$ and hence  more information than in the global version.

\end{document}

%% file: 20171117PRL_MT.bbl
%